\documentclass[11pt]{article}

\usepackage{authblk}
\usepackage{orcidlink}
\usepackage{hyperref}
\usepackage{fullpage}

\usepackage{amsmath}
\usepackage{booktabs}
\usepackage{enumitem}
\usepackage{subcaption}
\usepackage{tcolorbox}
\usepackage{graphicx}

\usepackage[
    backend=biber,
    style=numeric,
    sorting=none
]{biblatex}

\newcommand{\articletype}[1]{}
\newcommand{\email}[1]{%
  \par\noindent\textbf{Corresponding author: }
  \href{mailto:#1}{#1}\par
}
\newcommand{\keywords}[1]{%
  \par\noindent\textbf{Keywords: }#1\par
}

\newtcolorbox{codexprompt}{
  colback=gray!8,
  colframe=gray!45,
  boxrule=0.5pt,
  arc=2mm,
  left=2mm,
  right=2mm,
  top=1mm,
  bottom=1mm,
  fonttitle=\bfseries,
  title=Codex prompt
}

\graphicspath{{ioplatextemplate/}{figures/}}
\setlist[itemize]{leftmargin=*, itemsep=2pt, topsep=4pt}

\title{AI-Accelerated Gyrokinetic Predictions of Turbulent Transport for
Stellarator Design Optimization and Experimental Planning}

\author[1]{R. Michael Churchill\,\orcidlink{0000-0001-5711-746X}}
\author[2]{Matt Landreman}
\author[3]{Jong Youl Choi}
\author[2]{Byoungchan Jang}
\author[2]{Rory Conlin}
\author[4]{Noah Mandell}
\author[5]{Anima Anandkumar}
\author[5]{Valentin Duruisseaux}
\author[6]{Jeffrey Larson}
\author[7]{Dario Panici}
\author[7]{Yigit Gunsur Elmacioglu}
\author[1]{Jai Sachdev}
\author[8]{Tony Qian}

\affil[1]{Princeton Plasma Physics Laboratory, Princeton, NJ, United States of America}
\affil[2]{University of Maryland, College Park, MD, United States of America}
\affil[3]{Oak Ridge National Laboratory, Oak Ridge, TN, United States of America}
\affil[4]{Type One Energy, Knoxville, TN, United States of America}
\affil[5]{California Institute of Technology, Pasadena, CA, United States of America}
\affil[6]{Argonne National Laboratory, Lemont, IL, United States of America}
\affil[7]{Princeton University, Princeton, NJ, United States of America}
\affil[8]{University of Wisconsin, Madison, WI, United States of America}

\date{}

\begin{document}

\maketitle

\email{rchurchi@pppl.gov}


\begin{abstract}
Previous work built AI-based surrogates for a nonlinear gyrokinetic simulation code with the goal of using them for fast, direct calculations of turbulent ion heat flux in stellarator design optimizations and scenario planning for experiments. These AI surrogates were trained on data from \textgreater{}200k nonlinear, adiabatic electron gyrokinetic simulations with the gyrokinetic flux-tube code GX, using a wide range of stellarator magnetic configurations ($\sim$23k), positions in the plasma, and gradient scale lengths. In this paper, we demonstrate the use of the AI-based turbulence surrogate in the optimization of stellarator magnetic equilibrium and to speed up stellarator transport solvers. Due to its speed ($\sim$ms), the AI-based surrogate enables previously unattainable optimization objectives, such as full radial profiles of ion turbulent heat flux, or directly optimizing to maximize the turbulent critical gradient at multiple locations across the plasma. These direct calculations provide a potentially more accurate optimization target and reduce reliance on ad-hoc heuristics that may not accurately capture the variation of turbulent transport with magnetic configuration. By including the AI-based surrogate for turbulent heat flux in a transport solver, we can quickly postprocess and confirm the improved ion temperature resulting from the optimized equilibrium. Finally, we demonstrate the use of AI agents with strong reasoning AI models to automate the outer loop, exploring many objective and hyperparameter configurations with this AI-based turbulence surrogate to discover improved turbulence optimized magnetic equilibria.
\end{abstract}

\section{Introduction}
Stellarators are a promising path to a fusion power plant, but require careful design optimization of the magnetic coils to maximize fusion performance while meeting the many physics and engineering constraints. This is a high dimensional space to optimize in (thousands of degrees of freedom), with many local minima, suggesting the need for speed and robustness in the calculation of desired fusion machine metrics during the optimization loop. The natural and dominant path to achieve this has been to create physics-based proxies for complicated calculations, such as turbulent transport and fast ion loss, which can then be calculated quickly inside the optimization loop. While this route has been successful for a variety of metrics, there are complicated quantities where the physics-based proxy has actually been shown broadly to have little correlation with the direct calculation, for example the $\Gamma_c$ metric with fast ion loss.

Among the physics quantities of chief importance for predicting fusion performance is turbulent transport. In modern fusion machines, including now neoclassically-optimized stellarators such as W7-X, turbulence has been the dominant transport mechanism limiting fusion plasma performance. Yet optimizing stellarator magnetic coils to minimize turbulence is a difficult challenge, as the direct simulation of turbulent transport involves notoriously computationally expensive nonlinear gyrokinetic simulation codes. Here again a variety of physics-based proxies have been formed for turbulence transport for use in stellarator design optimization loops~\cite{mynick_optimizing_2010,kim_optimization_2024,jorge_direct_2023,stroteich_seeking_2022}.

In order to have a more direct calculation of the turbulent heat flux inside the stellarator design optimization loop, our previous work~\cite{landreman_how_2025} created AI-based surrogates of nonlinear gyrokinetic turbulence simulations generated by the \texttt{GX} code~\cite{mandell_gx_2024}. These AI-based surrogates were trained on over \textgreater{}200k nonlinear adiabatic electron gyrokinetic simulations, using a wide range of stellarator magnetic configurations (\textasciitilde{}23k), positions in the plasma, and gradient scale lengths. This previous work also showed the degree to which commonly used physics-based proxies capture the true turbulent ion heat flux from direct simulation calculation (see Figure 15 of~\cite{landreman_how_2025}, showing that AI-based surrogates have advantages in terms of generalizability to a wide variety of magnetic geometries). 

In this paper we demonstrate the utility of these AI-based turbulent ion heat flux surrogates by implementing them as an objective in the magnetic equilibrium optimizer DESC~\cite{dudt_desc_2020,panici_desc_2023,conlin_desc_2023, dudt_desc_2023} and in the transport solver T3D~\cite{qian_stellarator_2022}. We show in examples of increasing required number of turbulent ion heat flux evaluations the benefit of the combined speed and accuracy of the AI-based surrogate. These examples progress from simple single radial position calculations of turbulent ion heat flux, to tens of evaluations to optimize the turbulent critical gradient across the plasma, to hundreds of evaluations for the full time-dependent transport solves. These capabilities open possibilities for more accurate optimization and transport calculations, ultimately allowing for faster overall design iterations.

The paper is organized with the following sections: Section~\ref{sec:background} reviews previous efforts to include turbulence calculations in the optimization loop, and reviews in more detail the AI-based surrogate created, Section~\ref{sec:integration} describes the integration of the AI-based surrogate in the equilibrium optimizer \texttt{DESC} and the transport solver \texttt{T3D}, and demonstrate the advantage the AI-based surrogate brings, Section~\ref{sec:future} discusses current limitations and future work, and Section~\ref{sec:conclusions} summarizes and proposes future research directions.

\section{Background}\label{sec:background}

\subsection{GX AI Surrogate Model}
We begin with a brief overview of the GX AI surrogate model which we will use in this paper (for additional details, see Landreman et al.~\cite{landreman_how_2025}). As with any AI surrogate model, there are four important areas to describe: simulation code, data, architecture, training. 

\subsubsection{Simulation code}
The AI surrogate model is trained on data from the \texttt{GX} simulation code. \texttt{GX} solves the local $\delta f$ gyrokinetic equation in a field-aligned flux-tube domain using an Eulerian pseudo-spectral formulation, and supports both axisymmetric tokamak and fully three-dimensional stellarator magnetic geometries. The time-evolved perturbed plasma particle distribution function $\delta f(x,y,z,v_\parallel,\mu,t)$ is five-dimensional, with $z$ the coordinate along the magnetic field line, $x$ the radial coordinate normal to the flux surface ($\hat{z} \cdot \nabla x = 0$, with $\hat{z}$ the unit vector along the magnetic field), $y$ the binormal coordinate, $\hat{y} = \hat{b} \times \hat{x}$, $v_\parallel$ particle velocity parallel to the magnetic field, and $\mu = m v_\perp^2 / 2 B$ the perpendicular velocity component in conserved first adiabatic invariant (magnetic moment) form. A Fourier--Laguerre--Hermite pseudospectral formulation is used, using a Fourier basis for $x$ and $y$, a Hermite basis for $v_\parallel$, and a Laguerre basis for $\mu B$. Zonal flows ($k_y=0$) are retained in the simulations, and care taken not to introduce filtering applied to other $k_y>0$ modes since zonal flows are physically periodic and the spectral representation fulfills this boundary condition. 

\begin{figure}[bht]
    \centering
    \includegraphics[width=0.5\linewidth]{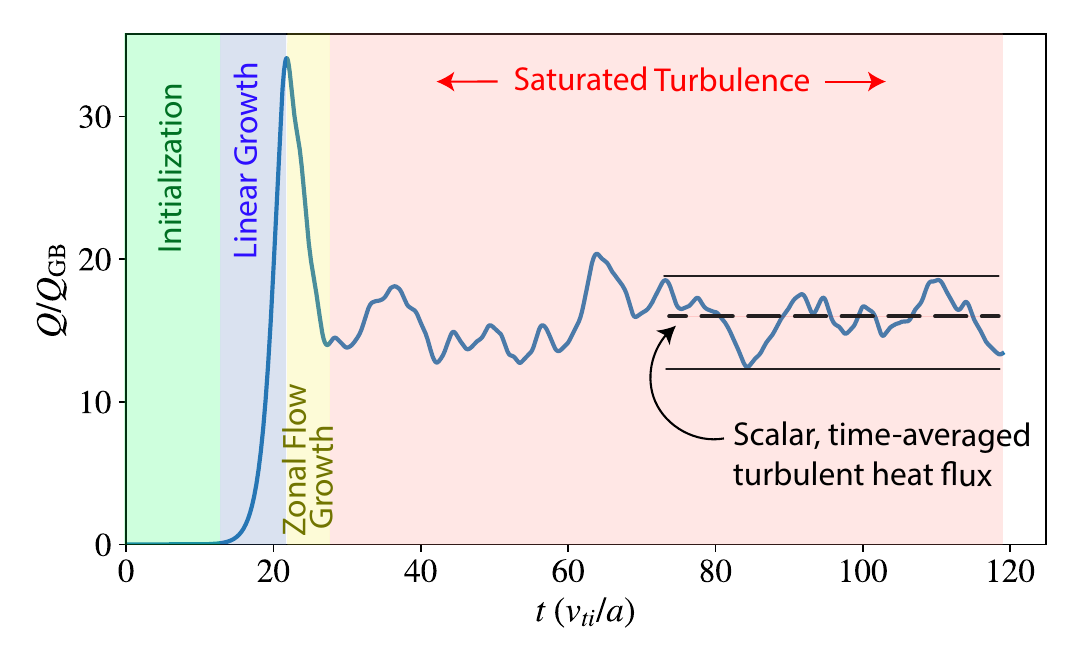}
    \caption{Evolution of the energy flux in data representative of a typical \texttt{GX} simulation. Illustrative/schematic example; not output from a specific GX simulation.}
    \label{fig:gx_sim}
\end{figure}

A representation of the typical simulation evolution of the normalized heat flux moment of the distribution function is shown in Figure~\ref{fig:gx_sim}. The system typically evolved in four stages: initialization with small-amplitude noise, to the unstable linear growth phase of an unstable mode (in the simulations for this paper the Ion Temperature Gradient (ITG) mode), followed by a nonlinear saturation mechanism (for example zonal flow growth which arrests the linear growth), and finally a quasi-steady stationary state of saturated turbulence. In addition to zonal-flow regulation, nonlinear saturation can involve energy transfer to subdominant or damped modes and other finite-$k_y$ interactions, with the relative importance of these channels depending on the three-dimensional magnetic geometry \cite{hegna_theory_2018,pueschel_stellarator_2016}. Typically a separation of scales is applied to transport equations for fusion plasmas~\cite{barnes_direct_2010}, where time-averaged turbulent fluxes are used as inputs to a longer-time transport solver. The quantity of interest then from these $\texttt{GX}$ simulations is the normalized time averaged energy flux, $\overline{Q}/Q_{GB}$.

In this paper we will focus on electrostatic, adiabatic electron simulations, with a local, flux-tube gyrokinetic code. The stellarator geometries we consider here are reactor size in major radius and field, which are generally considered to be ITG dominated due to the required large ion temperature. While this marginally may justify neglecting the effects kinetic electrons or finite-$\beta$ have on destabilizing new turbulence modes such as the trapped-electron mode (TEM) or Kinetic Ballooning Mode (KBM), these effects can also partially or fully stabilize or destabilize the ITG mode itself by modification of the adiabatic response \cite{xanthopoulos_gyrokinetic_2007,aleynikova_kinetic_2018}. Section \ref{sec:future} discuss future work including these effects. Additionally, we also note that local flux tube gyrokinetic codes can have differences from global calculations \cite{banon_navarro_global_2020}. While some works report favorable comparisons when ensuring sufficient length flux tubes\cite{smoniewski_comparison_2021,sanchez_gyrokinetic_2021}, it remains future work to determine if and how more detailed gyrokinetic calculations can be included in stellarator optimization problems. 

The \texttt{GX} code was designed specifically for speed, with the numerical algorithm and programming language (low-level \texttt{CUDA}) chosen to take advantage of the massive parallelization speedup of GPUs. This efficiency enables nonlinear adiabatic electron simulations in under 10 minutes on a single NVIDIA A100 GPU. This contrasts to hours for similar simulations on CPUs in the 2000's used by e.g., Mynick et al.~\cite{mynick_optimizing_2010} for turbulence stellarator optimization (see Section~\ref{sec:back_turb}).

\subsubsection{Data}
\texttt{GX} simulations were run for a dataset of approximately ~23.5k quasi-axisymmetric, quasi-helical stellarator, and random boundary configurations, which included partially optimized cases to improve robustness during optimization. For each equilibrium, four flux tubes were chosen, with a randomly sampled radial coordinate $\rho = \sqrt{s}$ ($s$ is the normalized toroidal magnetic flux) and with the field line label $\alpha = 0, \pi, -\iota \pi/N_{fp}, \iota \pi/N_{fp}$ respectively ($N_{fp}$ the number of field periods of the stellarator). For each flux tube, two separate sets of scalar gradient scale lengths were used, one set fixed at $a/L_{T_i} = 3$, $a/L_{n} = 0.9$, and one set randomly sampled ($a/L_{T_i}=\rho X_T$ with $X_T\sim\mathcal{N}(4,3^2)$ and a lower minimum of $X_T\ge 1.5$) to diversify data. 
No constraints on rotational transform were prescribed, beyond adjusting to make all configurations have positive $\iota$. Periodic boundary conditions were used in part to avoid increased resolution needs with the conventional twist-and-shift boundary conditions, and the varying flux tube length for the generalized twist-and-shift boundary conditions ~\cite{martin_parallel_2018, landreman_how_2025}. Chosen mesh size was $(N_{x}, N_{y}, N_z, N_{v_\parallel}, N_\mu) = (64,64,96,8,4)$. The scalar time-averaged heat flux was extracted from the end of the simulation, which was run for a total time of $t v_{th,i} / a = 800$. The transient behavior for $t v_{thi}/a_{minor} < 150$ was ignored, and a mean of the heat flux was computed over the remaining simulation time. Seven magnetic quantities used in the gyrokinetic equation were extracted from the equilibrium, as functions of $z$ along the \texttt{GX} mesh. These were $B,\; \mathbf{B}\times\nabla B\cdot\nabla x/B^3,\; \mathbf{B}\times\nabla B\cdot\nabla y/B^3,\; \mathbf{B}\times\boldsymbol{\kappa}\cdot\nabla y/B^2,\; |\nabla x|^2,\; \nabla x\cdot\nabla y,\; |\nabla y|^2$. Fixed flux tube lengths were used and same input sizes selected for the various simulations, so that the machine learning problem became mapping the input magnetic quantity array of size $[97, 7]$ and two gradient scale lengths to the output \texttt{GX} scalar ion heat flux $Q_i/Q_{i,GB}$.

\subsubsection{Architecture}
While a variety of AI models were trained in Landreman et al.~\cite{landreman_how_2025}, in this paper we will use only the convolutional neural network model, which achieved the best predictive performance. The architecture consists of one-dimensional convolutional layers typically with kernel size around $k = 20$ applied along the field-line coordinate $z$, with max pooling layers in between with average reduction factors of 2x. Global average pooling layer was used at the end. Periodic convolutions were used to enforce the exact periodicity of the GX flux-tube representation. Encoding this symmetry improves learning and generalization. 

\subsubsection{Training}
The DeepHyper framework~\cite{egele_deephyper_2025} is used for hyperparameter optimization, producing many trained variations of the CNN. The best 100 are kept, to form an ensemble at inference time, for improved robustness and to produce prediction uncertainty of the model. Data was randomly split in 80:20 for training:validation datasets. The output target data of the normalized ion heat flux $Q/Q_{GB}$ was limited with a floor of $10^{-2}$, to prevent the neural network from spending capacity on learning exact values of low heat flux levels.

The resulting trained ensemble achieves $R^2=0.989$ on the validation set (see Figure 9 in Landreman et al.~\cite{landreman_how_2025}). %

\subsection{Turbulent transport in stellarator optimization}\label{sec:back_turb}
Here we review various works which have developed physics-based proxies of turbulent transport for stellarator optimization. Mynick et al.~\cite{mynick_optimizing_2010} developed a physics based proxy for turbulent diffusivity from the Ion Temperature Gradient (ITG) mode, based on quasilinear estimates $\chi = \sum_k D_k$, with $D_k \propto \gamma / k_x^2$, where $\gamma$ is the linear growth rate and $k_x$ the radial component of the wave vector. Using a simplified ITG dispersion relation and mixed-length saturation, they derive equations for the growth rate and wave vector leaving four unknown physics parameters, which they fit using 12 nonlinear gyrokinetic \texttt{GENE} runs. They showed with the proxy they could optimize NCSX stellarator equilibrium to reduce turbulence transport, though the quasilinear proxy calculated orders of magnitude reduction, while post-optimization \texttt{GENE} simulations showed more modest 2-2.5x decrease. This suggests the proxy can be directionally correct for optimization purposes, but absolute reduction levels could not be relied on. Stroteich et al.~\cite{stroteich_seeking_2022} created a geometric proxy, targeting reduction of the flux expansion in regions of bad curvature, hypothesizing this should directly reduce the gradient drive for the ITG. They use a single flux tube centered at the midplane. They found a 50\% reduction in ion turbulent heat flux, and also that this is a result of reduced stiffness versus change in critical gradient location (``stiffness'' referring to the slope in the turbulent heat flux beyond the critical gradient, see Figure \ref{fig:critical_gradient}). Jorge et al.~\cite{jorge_direct_2023} took an important step in bringing the direct linear, adiabatic electron (e.g., ITG only) gyrokinetic simulation (\texttt{GS2}) calculation inside of the optimization loop, using a quasilinear estimate to derive the predicted heat flux. For speed sake, the optimization is performed for one field line on one flux surface, at one $k_x$ value and one set of gradient scale lengths. At this fixed gradient scale length, they do see a resulting 65\% reduction in turbulent heat flux (confirmed with nonlinear simulation done post-optimization). Finally, Kim et al.~\cite{kim_optimization_2024} recently brought the direct nonlinear gyrokinetic simulator ($\texttt{GX}$) inside the optimization loop, running adiabatic electron simulations for every optimization iteration to calculate ion heat flux from the ITG mode. They successfully reduced the turbulent heat flux by 2 - 4x, again by a reduction in stiffness, however due to time constraints of running the nonlinear gyrokinetic simulations in the optimization loop, they were limited to optimizing on only up to two flux tubes on a single flux surface ($s = 0.5$, $\alpha = 0, \iota \pi/4$). Even then, it took 10 minutes per optimization iteration, for a total of 12 hours for converged optimized equilibrium on a single NVIDIA A100 GPU.   

These works showed the promise of including turbulent heat flux as an optimization target, yet all had to make simplifications in terms of the fidelity of the turbulent heat flux calculation. Indeed, in Figure 15 of Landreman et al.~\cite{landreman_how_2025} compares many of the traditional ITG heat flux proxies, including those derived from the AI models trained therein, showing many of these proxies correlate at various levels with the ion heat flux, but overall $R^2$ performance lags behind the best trained model (the CNN).

\section{Integration with Plasma Transport and Equilibrium Optimization Workflows} \label{sec:integration}
We turn now to the application of the AI \texttt{GX} model in stellarator equilibrium optimization and transport solvers.

\subsection{Equilibrium Optimization}
For stellarator magnetic equilibria optimization we use the JAX-based \texttt{DESC} code~\cite{dudt_desc_2020}, and use the AI \texttt{GX} turbulence surrogate in various objectives, as detailed in the following. A key benefit of the neural network based surrogate compared to the full \texttt{GX} simulation in the optimization loop is that the neural network is by nature differentiable, and so avoids multiple \texttt{GX} runs needed for gradient calculations using finite differencing or stochastic gradient approximations~\cite{kim_optimization_2024}. For all of these cases, we initialize the optimization with the Landreman-Paul precise QA stellarator equilibrium~\cite{landreman_magnetic_2022}, adjusted for reactor scale and found as an example in the \texttt{DESC} code repo. 

\subsubsection{Turbulence optimization - Single flux tube}
First, we optimize the equilibrium using as an optimization target the surrogate heat flux prediction at a single flux tube at $s = 0.5$ (similar to Kim et al.~\cite{kim_optimization_2024}). Fixed values of $a/L_{n}=1.0$ and $a/L_{T_i}=3$ are used during the optimization. With the AI \texttt{GX} surrogate, this optimization took less than 10 minutes on a single A100 GPU, roughly a factor of 72x faster than when using the actual \texttt{GX} simulation in the loop. For this optimization, for the objective function we target the two-term quasisymmetry error, the ion heat flux at $s = 0.5$ and a term to force the aspect ratio to stay the same as the initial equilibrium:

\begin{equation}
    f = \lambda_{QS} f_{QS}^2 + \lambda_{Q^{GX}} f^2_{Q^{GX}(s = 0.5, \alpha = 0)}+ \lambda_A (A - A_{target})^2
    \label{eq:phase0_objective}
\end{equation} 

where the quasisymmetry two term objective~\cite{helander_theory_2014} $f_{QS} = \sum_{s_j} \langle 
\left( \left[\iota (\mathbf{B} \times \nabla\psi)- G\mathbf{B}\right]\cdot \nabla B / B_0^3\right) \rangle$ for quasi-axisymmetry, with $G = \mu_0/(2\pi)$ times the poloidal current outside the surface. The $\lambda$ factors are independent scabale weighting factors. At each optimization step, Fourier coefficients describing the boundary shape are modified, after which \texttt{DESC} re-solves the ideal MHD equilibrium by minimizing the force-balance residual, $\mathbf{F} = \mathbf{J} \times \mathbf{B} - \nabla p$. The resulting new equilibrium is represented by Fourier-Zernike spectral basis coefficients $R_{lmn}$, $Z_{lmn}$, and $\lambda_{lmn}$, where $R_{lmn}$ and $Z_{lmn}$ describe the geometry of the nested flux surfaces in cylindrical coordinates and $\lambda_{lmn}$ describes the poloidal stream function relating the computational poloidal coordinate to the straight-field line coordinate. $l$ is the radial Zernike basis, $m$ the poloidal mode number, and $n$ the toroidal mode number. The current profile, pressure profile, and toroidal flux are held fixed, while the rotational transform $\iota$ is allowed to vary. The AI GX surrogate then calculates the heat flux using the new equilibrium, and the new objective function (Equation \ref{eq:phase0_objective}) is evaluated.

\begin{figure}[bht]
    \centering
    \begin{subfigure}[t]{0.48\linewidth}
        \centering
        \includegraphics[width=\linewidth]{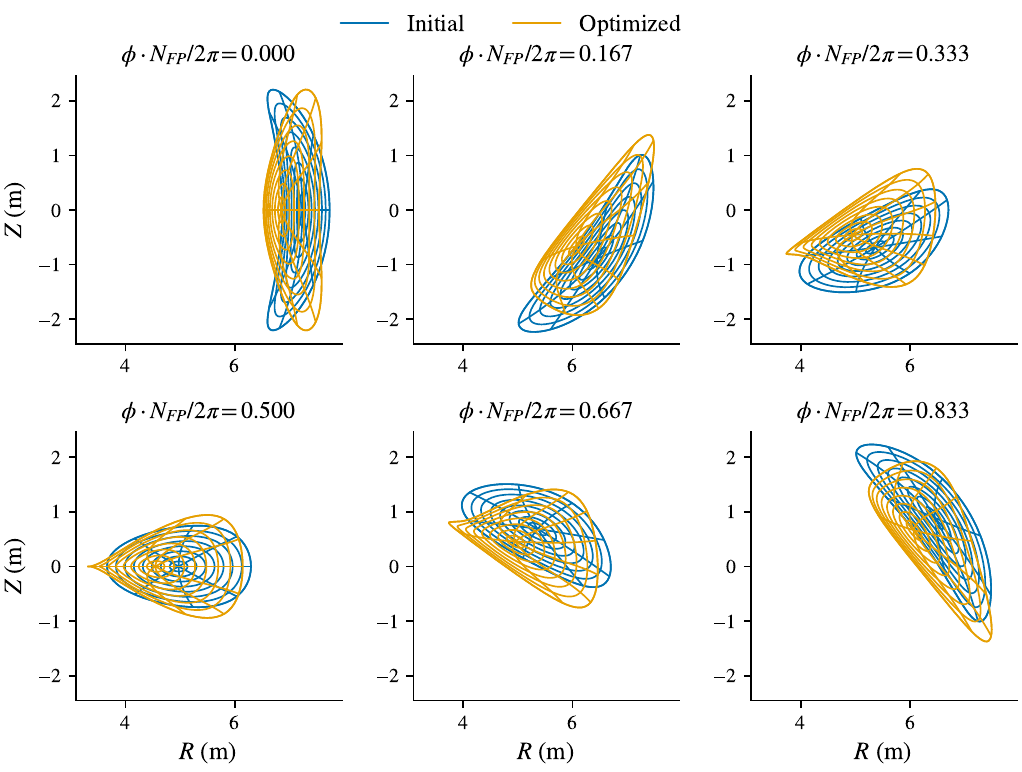}
        \caption{Initial and optimized equilibrium cross-sections at several toroidal angles.}
        \label{fig:phase0_equilibrium}
    \end{subfigure}
    \hfill
    \begin{subfigure}[t]{0.48\linewidth}
        \centering
        \includegraphics[width=\linewidth]{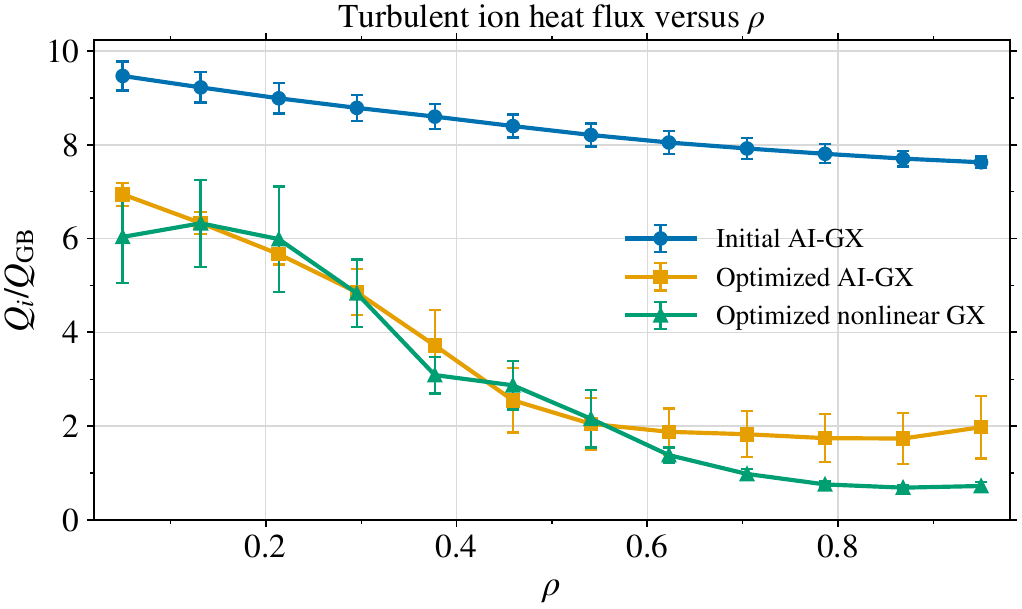}
        \caption{Initial and optimized normalized ion heat flux versus radial coordinate $\rho$}
        \label{fig:phase0_Q_vs_rho}
    \end{subfigure}

    \caption{Results from equilibrium optimization using AI \texttt{GX} surrogate ion heat flux prediction at a single flux tube}
    \label{fig:phase0_opt}
\end{figure}

The initial and final equilibrium and ion heat fluxes are shown in Figure~\ref{fig:phase0_opt}. As seen, the optimization is successful in reducing the ion heat flux not only at $\rho = \sqrt{s} = \sqrt{0.5} \approx 0.7$, but also across the board, ranging from a 25--50\% reduction. The effective helical ripple $\varepsilon_{eff}$, a measure of neoclassical transport caused by departures from perfect quasisymmetry, does increase significantly, in the initial equilibrium ranging from $3.8 \times 10^{-7} - 4.9 \times 10^{-6}$ and in the final equilibrium $9.2 \times 10^{-4} - 2.4 \times 10^{-2}$. While significant, we have not optimized for neoclassical transport, and it is still within the range of the W7-X design point of $\varepsilon_{eff} \approx 1\%$.

\subsubsection{Turbulence optimization - Multiple flux tubes}\label{sec:opt_multiple}
Next, we expand to an objective with the ion heat flux calculated at many flux tubes spanning $\rho$, again using midplane flux tubes ($\alpha = 0$), and $a/L_{n}=1.0$ and $a/L_{T_i}=3$. 

\begin{equation}
    f = \lambda_{QS} f_{QS}^2 + \lambda_{Q^{GX}} \sum_j f^2_{Q^{GX} (s = s_j, \alpha = 0)} + \lambda_A (A - A_{target})^2
\end{equation} 

Results are shown in Figure~\ref{fig:phase0_vs_phase2}, which shows the original single flux tube optimization (from Figure~\ref{fig:phase0_opt}) on the left, versus that of optimizing using multiple flux tubes on the right, both shown versus $\rho$ at $\alpha = 0$ as before on the top, and on the bottom versus field line label $\alpha$ at the $\rho \approx 0.7$ flux surface. Turbulent heat flux drops across the radius as before, but not as much as when using a single flux surface. However, viewing the change in ion heat flux versus $\alpha$ for the $\rho \approx 0.7$ flux surface, we can see variability. We postprocess to evaluate the heat flux on these equilibria at 8 $\alpha$ values per flux surface, performing the flux surface average as:

\begin{figure}
\begin{subfigure}[t]{0.48\linewidth}
    \centering
    \includegraphics[width=\linewidth]{figures/phase0_fixed_gradient_heat_flux.pdf}
    
\end{subfigure}
\hfill
\begin{subfigure}[t]{0.48\linewidth}
    \centering
    \includegraphics[width=\linewidth]{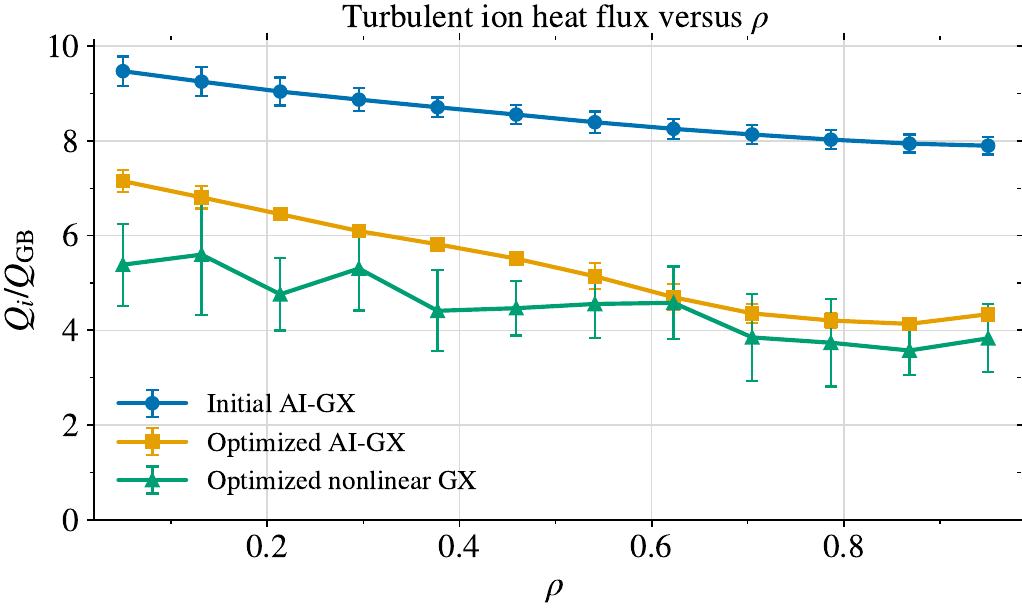}
\end{subfigure}

\begin{subfigure}[t]{0.48\linewidth}
    \centering
    \includegraphics[width=\linewidth]{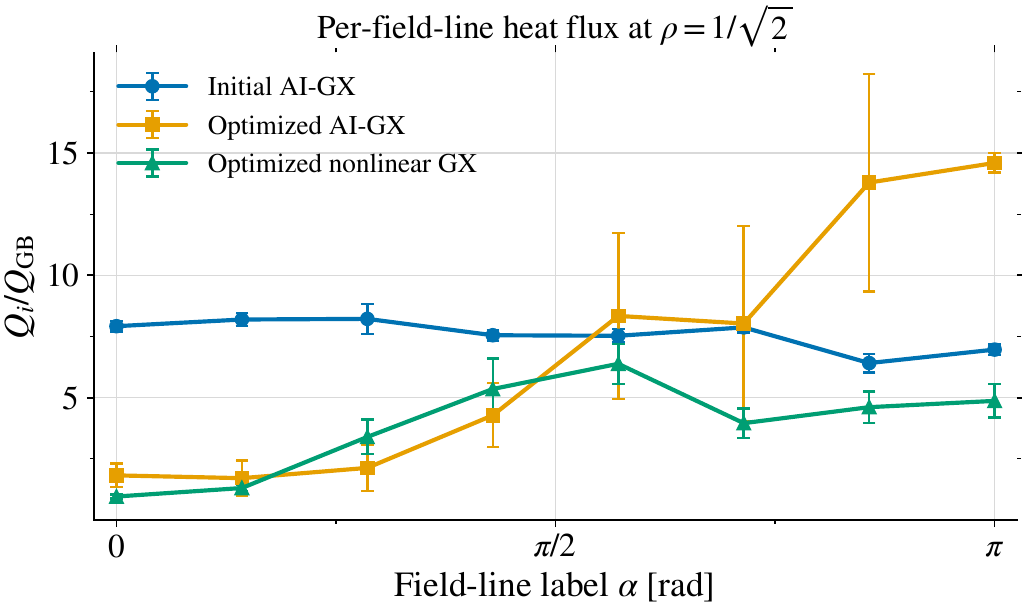}
\end{subfigure}
\hfill
\begin{subfigure}[t]{0.48\linewidth}
    \centering
    \includegraphics[width=\linewidth]{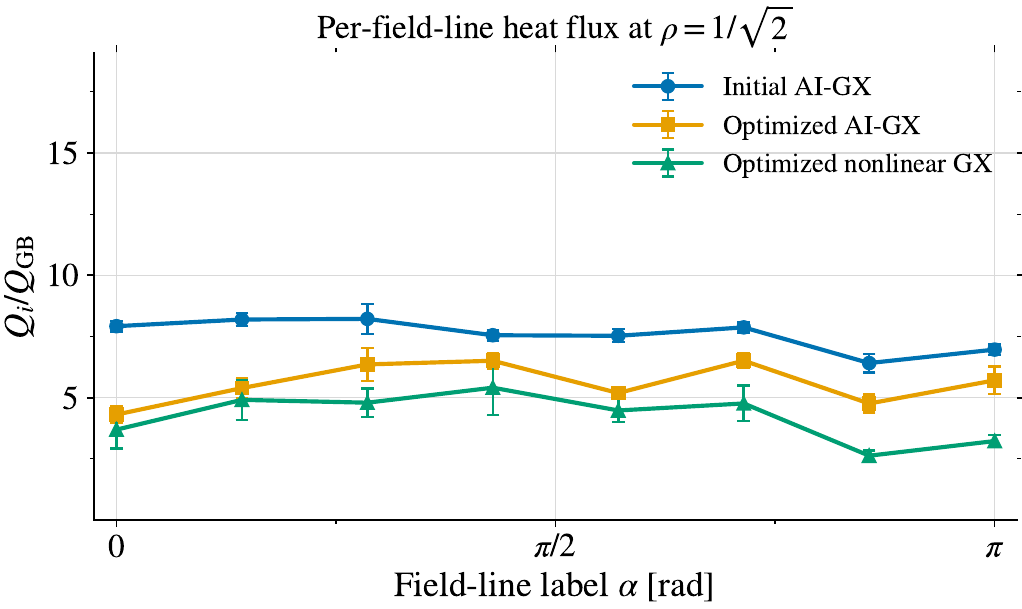}
\end{subfigure}

\caption{Results from equilibrium optimization using AI \texttt{GX} surrogate ion heat flux prediction at a single flux tube (left) and multiple flux tubes (right). The top row depicts $Q/Q_{\mathrm{GB}}$ versus $\rho$, and the bottom row versus $\alpha$ on the $\rho = \sqrt{0.5}$ flux surface. This demonstrates optimizing with multiple points can lower overall turbulence.}

\label{fig:phase0_vs_phase2}
\end{figure}

\begin{equation}
\left\langle Q \right\rangle(\rho)
=
\frac{1}{2\pi}
\int_{0}^{2\pi}
Q^{GX}(\rho,\alpha)\,d\alpha
\approx
\frac{1}{N_\alpha}
\sum_{k=1}^{N_\alpha}
Q^{GX}\left(\rho,\alpha_k\right),
\qquad
\alpha_k=\frac{2\pi k}{N_\alpha}.
\end{equation}

\begin{figure}%
    \centering
    \includegraphics[width=0.9\linewidth]{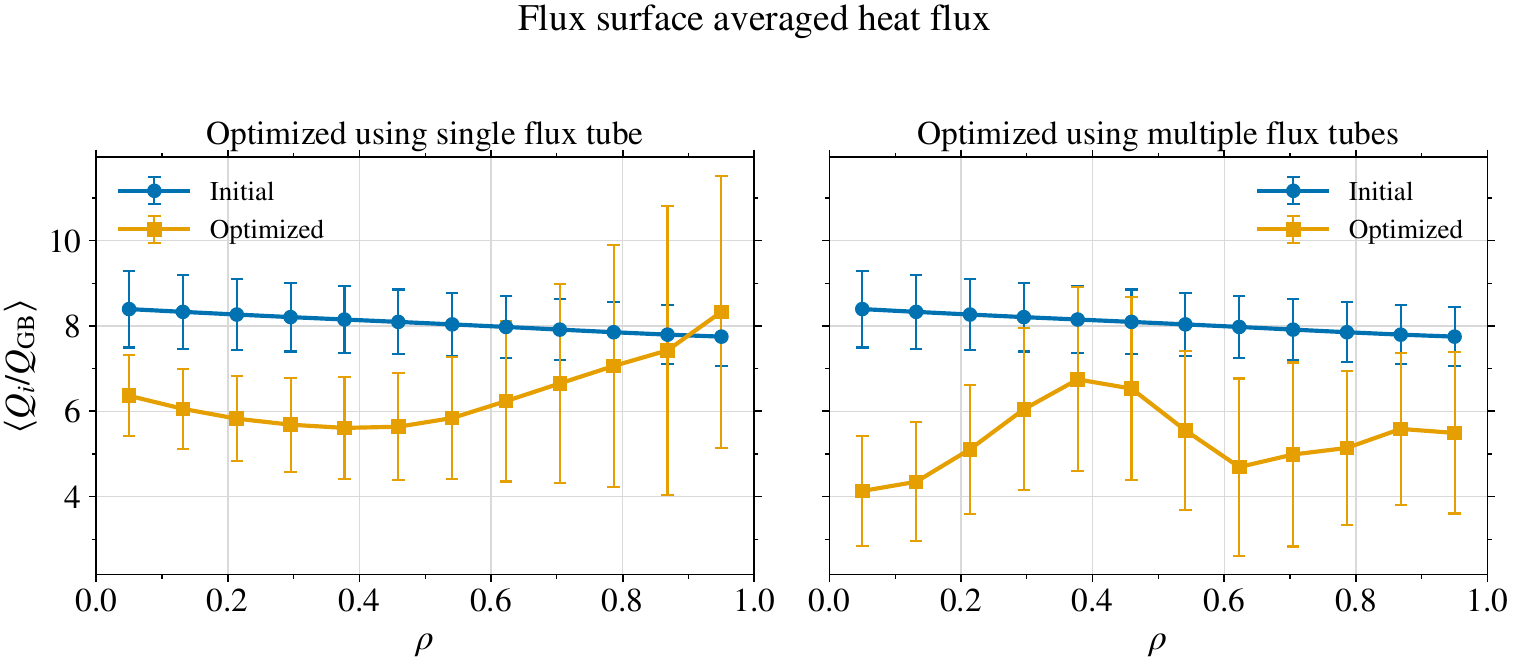}
    \caption{Flux-surface averaged heat flux using multiple field lines per flux surface, for the single flux tube optimization (left) and the multiple flux tube optimization (right)}
    \label{fig:phase2_fluxavg}
\end{figure}

(This direct sum over $\alpha$ is possible because the \texttt{GX} variable for heat flux already includes parallel integration including the $z$-coordinate Jacobian). The resulting flux-surface averaged heat flux from the single flux tube optimization and the multiple-flux tube optimization is shown in Figure~\ref{fig:phase2_fluxavg}, showing improvement across the radial profile by using multiple flux surfaces in the optimization.

\subsubsection{Turbulence optimization - Critical Gradient}
These results look promising for the ability to directly reduce the turbulent transport using the AI \texttt{GX} surrogate. However, when running these equilibria through the \texttt{T3D} transport solver (see Section~\ref{sec:t3d}), the resulting transport solution yields a very small heat flux improvement compared to what is shown in e.g., Figure~\ref{fig:phase0_opt} (as was seen also in Kim et al.~\cite{kim_optimization_2024}). It was realized that the above results are obtained at \emph{fixed} gradient scale lengths. Due to the stiff transport nature of turbulence in stellarator plasmas, typically the plasma profiles (e.g., $n$, $T$) develop up to a critical gradient where turbulence increases rapidly, keeping the profiles ``clamped'' near those levels. While the optimizations just shown reduce the turbulent heat flux levels, they do so at regions of gradient scale length space which the plasma will not evolve to. Reducing the ``stiffness'' (effectively what we do above) does not guarantee we will actually improve the metric we really care about, which is the ion temperature. 

\begin{figure}
    \centering
    \includegraphics[width=0.45\linewidth]{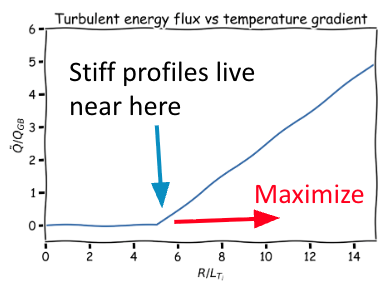}
    \caption{Idealized depiction of the goal of critical gradient maximization. The graph is of the turbulent heat flux versus gradient scale length, indicating stiff transport behavior for stellarators. The point where the heat flux begins increasing is the ``critical gradient'', and the slope of the heat flux beyond that point is referred to as the ``stiffness''}
    \label{fig:critical_gradient}
\end{figure}

Ideally, it would be preferred to include a full transport solve for the ion temperature, which could be a future work if the differentiability for the transport solver is worked out, for example extending a code like TORAX~\cite{citrin_torax_2026}, or by using optimization methods designed for settings where only some derivatives of the objective are available~\cite{LMT2026}.
Another option that is more immediately achievable is to optimize a differentiable proxy that drives the critical gradient to larger values at multiple radial and field-line locations (see Figure~\ref{fig:critical_gradient}). This now requires not just predicting the heat flux across space, but also scanning the gradient scale length $a/L_{T_i}$ ($a/L_n$ is held constant at 0.3) for each flux tube during each optimization iteration, to extract the location of the critical gradient, dramatically increasing the number of turbulence evaluations. As maximums can be problematic for differentiability, we use a smooth proxy:

\begin{equation}
\Phi(Q)
=
\Delta_{\ln(Q)} \operatorname{softplus}\!\left(F(Q)\right)
=
\Delta_{\ln(Q)}  \ln \!\left[
1+
\exp\!\left(
\frac{
\ln\!\left(\max\!\left(Q,Q_{\min}\right)\right)
-
\ln\!\left(Q_{\mathrm{crit}}\right)
}{
\Delta_{\ln(Q)}
}
\right)
\right].
\end{equation}

with the following choices: $Q_{min} = 1e-2$, the threshold to mark the critical gradient $Q_{crit} = 1$, and $\Delta_{\ln(Q)} = 0.35$. As before we choose 9 $\rho$ positions to evaluate heat flux at, keep $\alpha = 0$, and scan 8 $a/L_{T_i}$ values ranging from 0.25 to 4.25, giving 72 total AI \texttt{GX} surrogate evaluations per optimization iteration. These optimization still take $\approx 10$ minutes, whereas the full \texttt{GX} simulation code would require an enormous amount of time or additional GPUs to run. The critical-gradient optimization used DESC's \texttt{proximal-lsq-exact} optimizer, with objective-, step-, and gradient-based convergence tolerances all set to $10^{-4}$ and a maximum of 1000 iterations. As before, the total MHD pressure profile, the current, and the toroidal flux are held fixed, and the density gradient scale length $a/L_n$ at each position is also held fixed. The initial optimization showed good results, but had a number of sharp features in the equilibrium. These can be detrimental from the perspective of coil complexity required to create the equilibrium, and potential non-realizability when more faithful MHD and transport is included. To avoid this, two \texttt{DESC} geometric objectives were added, \texttt{PrincipalCurvature} which penalizes when the curvature exceeds specified bounds to avoid cusps, and GoodCoordinates which uses an algorithm to avoid self-intersecting curves. The resulting optimization target is:

\begin{equation}
f
=
\lambda_{\rm QS} f_{\rm QS}^{2}
+
\lambda_{\rm CG}\sum_j f_{\rm CG}^{2}(s=s_j,\alpha=0)
+
\lambda_A(A-A_{\rm target})^{2}
+
\lambda_{\kappa} f_{\kappa}^{2}
+
\lambda_{\rm GC} f_{\rm GC}^{2}.
\end{equation}

where $f_{QS}$ and $A$ are the same as in Equation \ref{eq:phase0_objective}, $f_{CG}$ is the critical gradient metric:

\begin{equation}
f_{\rm CG}(s_j,\alpha=0)
=
\frac{
\displaystyle
\sum_{\ell} w_\ell\,
\Phi\!\left[
Q_{\rm GX}(s_j,\alpha=0,g_\ell)
\right]
}{
\displaystyle
\sum_{\ell} w_\ell
},
\end{equation}

with the weights $w_\ell = (g_{max} - g_\ell) / (g_{max} - g_{min})$, where $g_\ell = \left(a / L_T\right)$, which encourage higher critical gradient. The \texttt{PrincipalCurvature} $f_\kappa$ regularizes the geometric principal curvatures of the plasma boundary surface: 

\begin{equation}
f_\kappa
=
\left\|
\max(\kappa_{\max}-\kappa_{\rm cap},0)
\right\|_2
\end{equation}

with $\kappa_{\rm max}$ the maximum geometric principle curvature of the surface, and $\kappa_{\rm cap} = 8 \text m^{-1}$. The \texttt{GoodCoordinates} regularization term $f_{GC}$ is used to discourage coordinate singularities/self-intersections.

\begin{figure}[htbp]
    \centering
    \begin{subfigure}[t]{0.48\linewidth}
        \centering
        \includegraphics[width=\linewidth]{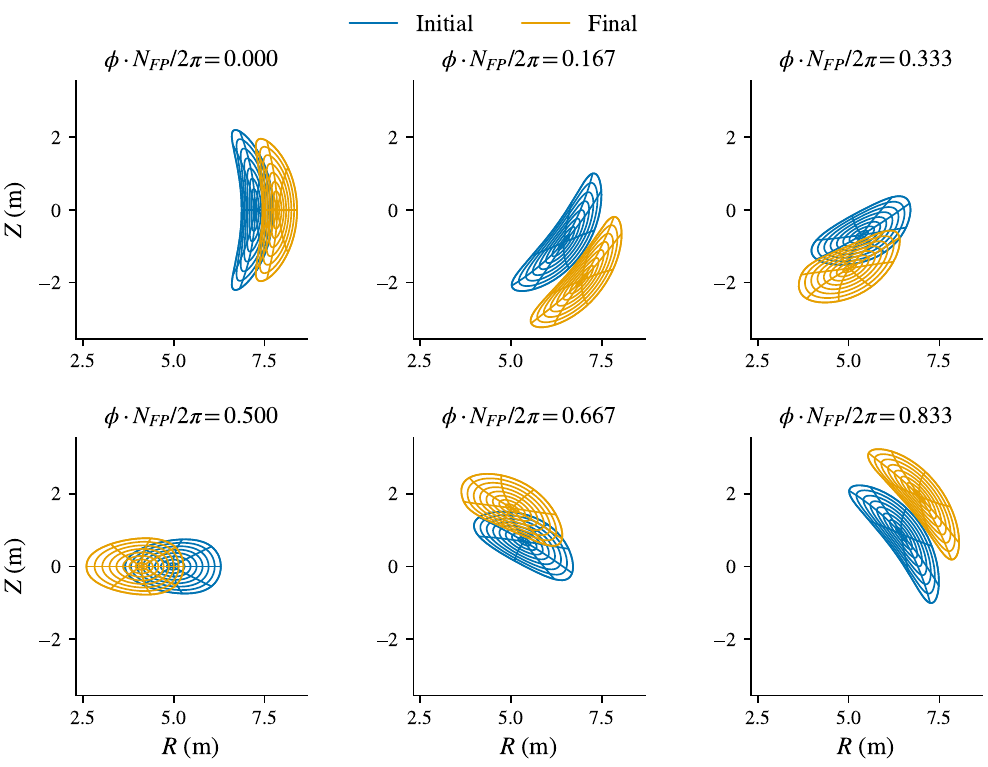}
    \end{subfigure}
    \hfill
    \begin{subfigure}[t]{0.48\linewidth}
        \centering
        \includegraphics[width=\linewidth]{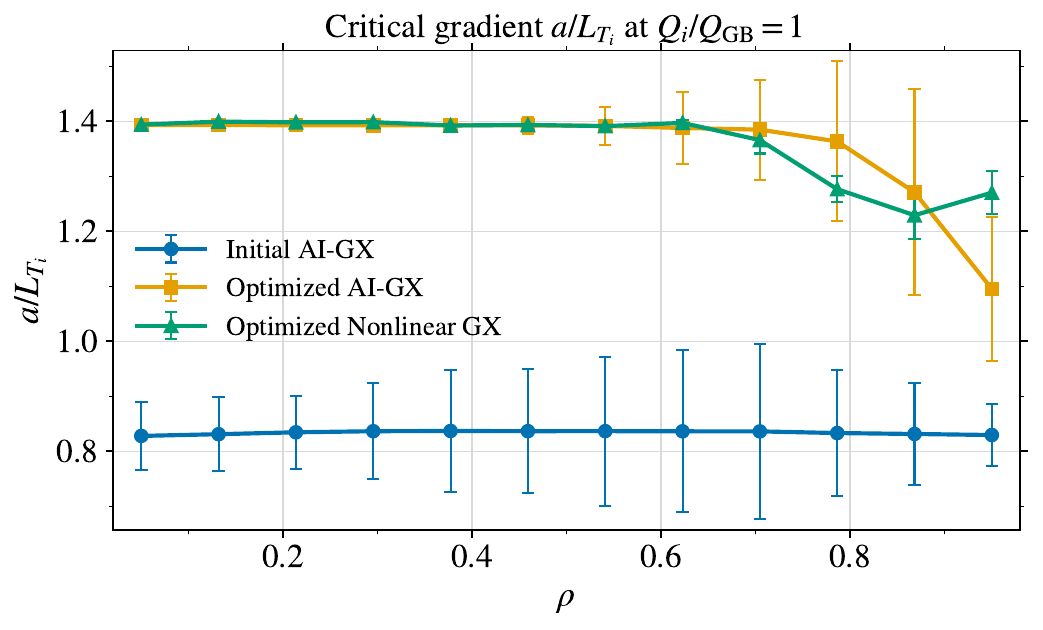}
    \end{subfigure}
    \caption{Results from equilibrium optimization using AI \texttt{GX} surrogate to optimize the critical gradient at each flux surface, (a) the initial and optimized equilibrium, (b) critical gradient (defined as gradient scale length $a/L_{T_i}$ where $Q_i/Q_{GB}=1$) versus radial coordinate $\rho$ for initial and optimized equilibrium. Final legend entry is a validation using the full nonlinear \texttt{GX} simulator on the optimized equilibrium.}
    \label{fig:phase3_results}
\end{figure}

The results of the optimization are shown in Figure~\ref{fig:phase3_results}, showing the resulting equilibrium cross-sections, and the critical gradient which improves varying from 35 - 70\% increase from the initial. The equilibrium shows smooth contours, and interestingly the optimization has resulted in an equilibrium with cross sections very similar to the initial equilibrium, but pushed outwards, leading to an increased torsion of the magnetic axis. 

\subsubsection{AI agents for Turbulence optimization}\label{sec:opt_agent}
Many of these optimizations in \texttt{DESC} require decisions on objectives to include, weighting of different objectives and constraints, hyperparameters for the constraints (e.g., number of flux surfaces to calculate heat flux on). The permutation space is rather large. To help explore the space more efficiently, we leverage AI agentic capability through the OpenAI Codex AI agent environment. We setup a problem similar to the \texttt{autoresearch} repository by Karpathy~\cite{karpathy_autoresearch_nodate}, where they provided natural language instructions to an AI agent powered by strong reasoning Large Language Models (LLMs). These instructions were to do a training of NanoGPT, modifying the code to try new architectural changes, modifying hyperparameters such as learning rate, etc. Note this is similar to Bayesian optimization methods for hyperparameter tuning, but much more expansive, as the reasoning LLM model can determine changes to make and edit code directly. Applying this to our \texttt{DESC} equilibrium optimization problem, we wanted to target improving upon the earlier optimization. The Codex AI agent has the capability to do long running, goal oriented tasks, specified by the \texttt{/goal} mechanism. We prompted Codex to create a detailed plan (see Appendix \ref{sec:appendix_codex_prompts}) which was then run in this persistent goal mode. Some minor edits were required to hold fixed inputs like the flux tube length to the AI \texttt{GX} surrogate, but beyond these no human intervention was needed besides extending the runtime a few times. This goal plan developed by Codex setup a genetic algorithm to perform this optimization (we note here we attempted subsequent AI agent optimizations instructing to not used fixed genetic algorithms, but rather the strong reasoning LLM models to determine next steps, but with the prompt given it seemed to get stuck in small incremental changes versus exploring a wider variety of optimization combinations, and so was left for future work). Codex setup the code to launch the simulations, the genetic algorithm to decide which changes to make in successive generations based on prior parent generations, and an optimization score metric, defined as:

\begin{subequations}
\label{eq:autoresearch-score}
\begin{align}
S(r)
&=
S_{\mathrm{turb}}(r)
+ 0.25\,f_{\mathrm{QS},r}^{\,2}
+ 0.25\,\left|A_r-A_r^\star\right|
+ 0.05\,\frac{\tau_r}{1~\mathrm{h}}
+ P_{\mathrm{stages}}(r),
\\[3pt]
S_{\mathrm{turb}}(r)
&=
\begin{cases}
\displaystyle
\log\!\left[
1+\max\!\left(\overline{Q}_{\mathrm{GX},r},0\right)
\right],
& \text{fixed-gradient optimization}, \\[8pt]
\displaystyle
-C_r+\left[C^\star-C_r\right]_+,
& \text{critical-gradient optimization},
\end{cases}
\\[3pt]
\overline{Q}_{\mathrm{GX},r}
&=
\frac{1}{N_\rho N_\alpha}
\sum_{j=1}^{N_\rho}
\sum_{k=1}^{N_\alpha}
Q_{\mathrm{GX},r}(\rho_j,\alpha_k).
\end{align}
\end{subequations}

Note that this is not a gradient-based optimization target, but rather a score metric assigned post run, used by the genetic algorithm to determine next changes. Here \(\overline{Q}_{\mathrm{GX},r}\) is the mean normalized turbulent ion heat flux obtained in the post-processing transport calculation for run \(r\). For critical-gradient runs, \(C_r\) is the critical value of \(a/L_{T_i}\), averaged over the sampled radial locations, and \(C^\star=3\) is the target used by default. The positive-part term \([C^\star-C_r]_+\) imposes an additional penalty when the achieved critical gradient falls below this target. The remaining terms penalize the quasisymmetry residual, deviation of the achieved aspect ratio \(A_r\) from its run-specific target \(A_r^\star\), runtime in hours, and $P_{stages}$ a penalty of $10^5$ if incomplete Fourier continuation (e.g., if a second round optimization for higher-order modes fails due to automated geometry extraction. In practice this penalty never occurred.). Runs that fail outright (dominantly because the automated geometry extractor failed to create a flux tube of the right length) are given a penalty of $10^6$. This objective is itself a composite function: a simple, known combination of quantities extracted from an expensive \texttt{DESC} run; we are actively considering structure-exploiting derivative-free optimization methods designed for exactly this composite setting~\cite{Larson2022} as a alternative or complement to the genetic algorithm initially considered.

The custom based genetic algorithm used generation-based evolutionary search with a population of 16 candidate optimizations per generation, with four evaluations performed concurrently on separate GPUs. After each generation, the four highest-ranked candidates were retained as elites, and the remaining population was generated through mutation and crossover of the elite configurations. Crossover was applied with probability 0.35, with individual parameters inherited independently from a second parent, followed by mutation in which up to four optimization parameters were resampled from their prescribed search distributions. The search varied parameters including the Fourier-mode continuation schedule, quasisymmetry and turbulence-objective weights, aspect-ratio target and weight, optimizer trust radius and convergence tolerances, and the radial and field-line sampling used for evaluation. 

\begin{figure}
    \centering
    \includegraphics[width=0.75\linewidth]{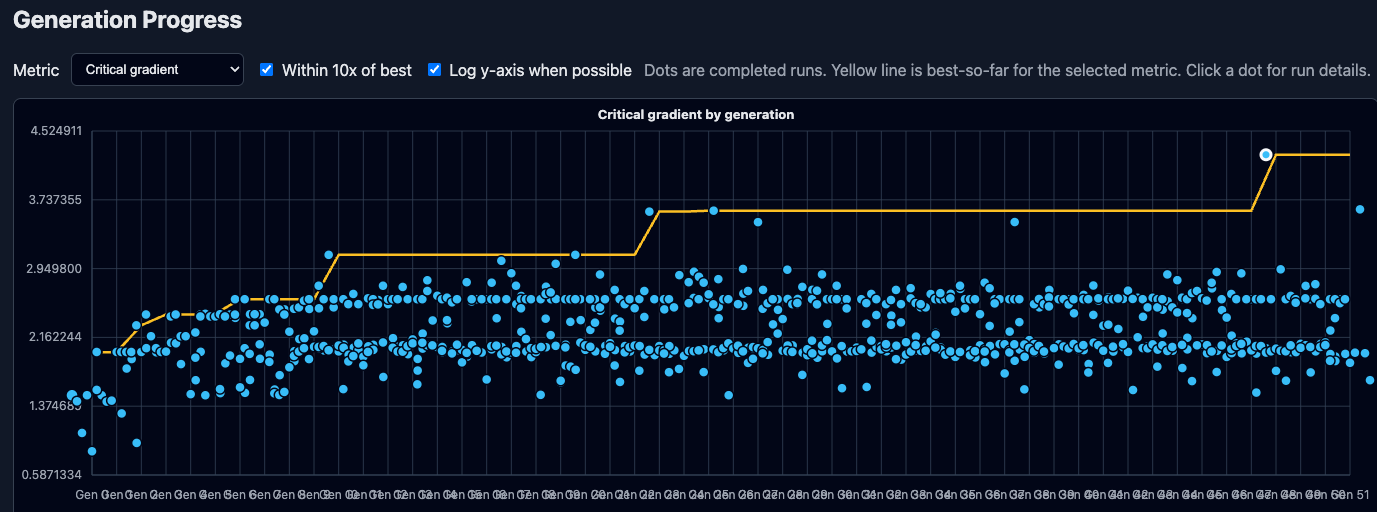}
    \caption{Results from the optimization campaign, which each dot representing the averaged critical gradient for the different \texttt{DESC} optimizations versus the generation of the genetic algorithm. }
    \label{fig:opt_campaign}
\end{figure}

We show the results of this optimization campaign\footnote{\href{https://ti-optimization-autoresearch.rmc256.chatgpt.site/}{https://ti-optimization-autoresearch.rmc256.chatgpt.site/}} in Figure~\ref{fig:opt_campaign}, with each dot representing a successful \texttt{DESC} optimization, with the x-axis showing the generations of the genetic algorithm and the y-axis the mean critical gradient. The outer search ran a total of 52 generations, for a total of 758 completed, 40 failed, 29 skipped, and 5 unfinished \texttt{DESC} optimizations. Individual \texttt{DESC} optimizations terminated according to their specified convergence tolerances or maximum iteration count. This took around a total of 72 hours, run in different days on a single Perlmutter compute nodes with four A100 GPUs. 

\begin{figure}[htbp]
    \centering
    \begin{subfigure}[t]{0.48\linewidth}
        \centering
        \includegraphics[width=\linewidth]{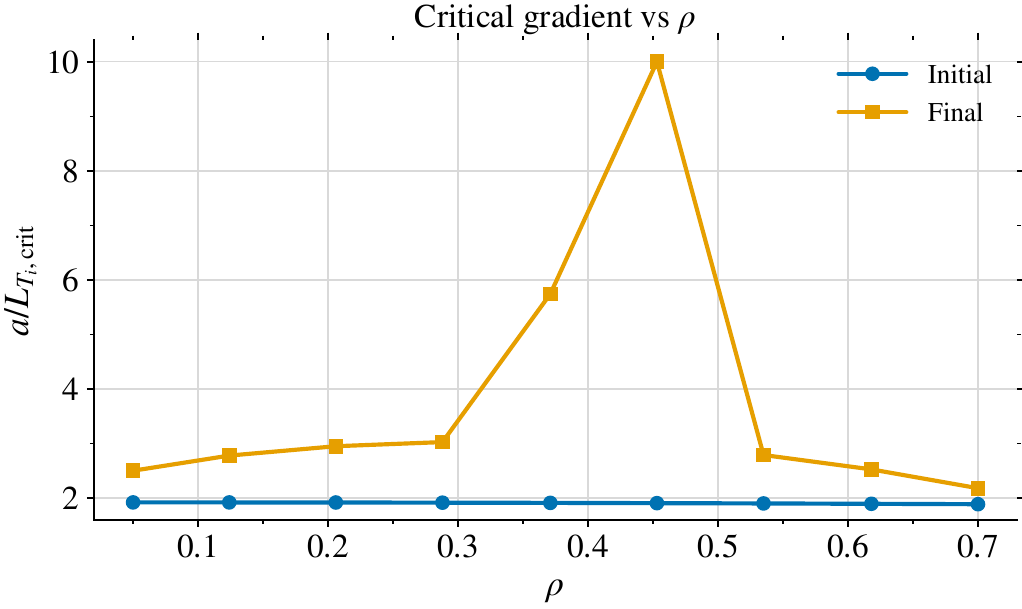}
    \end{subfigure}
    \hfill
    \begin{subfigure}[t]{0.48\linewidth}
        \centering
        \includegraphics[width=\linewidth]{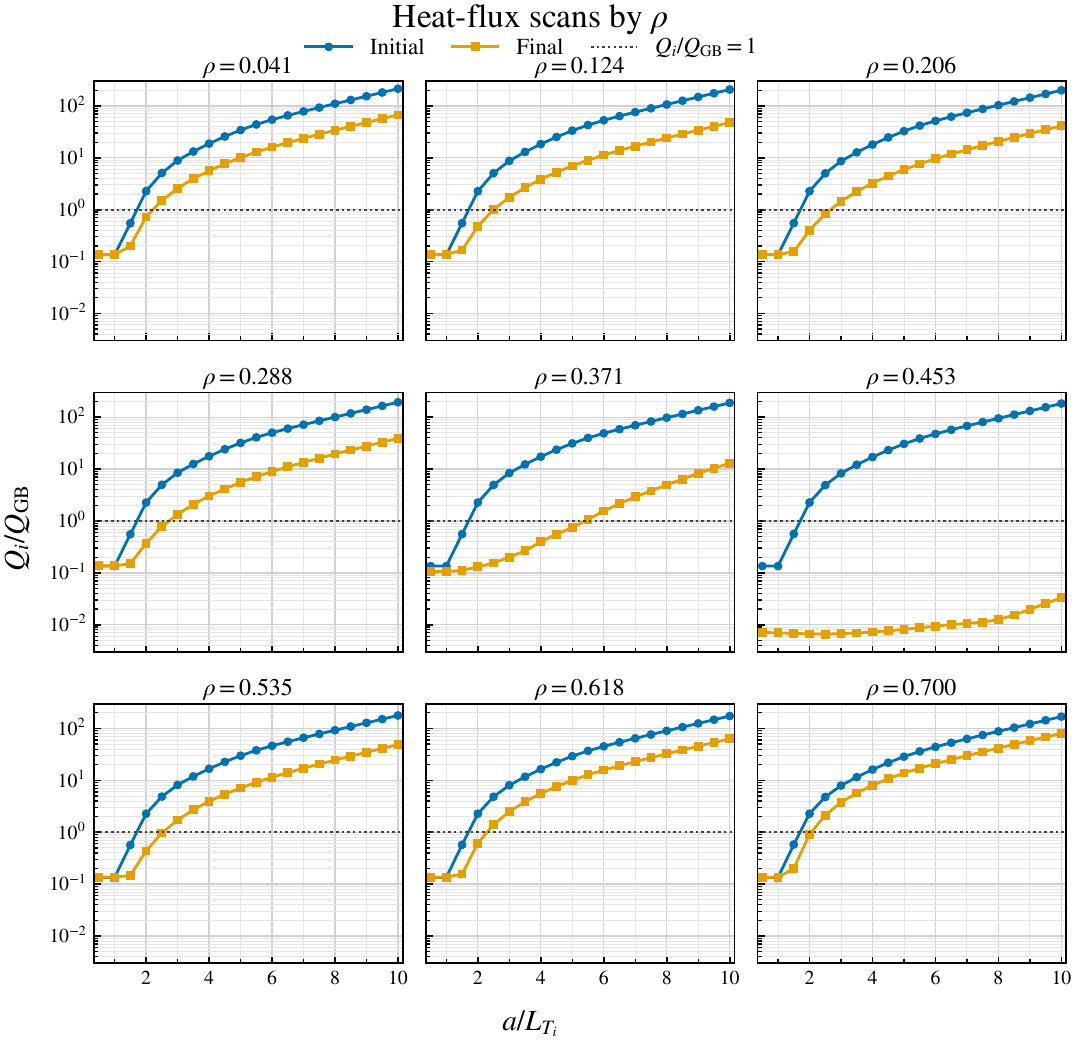}
    \end{subfigure}

    \caption{Best performing equilibrium from the Codex optimization. (a) Critical gradient versus $\rho$, showing the jump for two mid-radius points. (b) Heat flux versus $a/L_{T_i}$, showing these two mid-radius points have suspiciously low heat flux everywhere.}
    \label{fig:bad_critical}
\end{figure}

\begin{figure}
    \centering
    \includegraphics[width=0.75\linewidth]{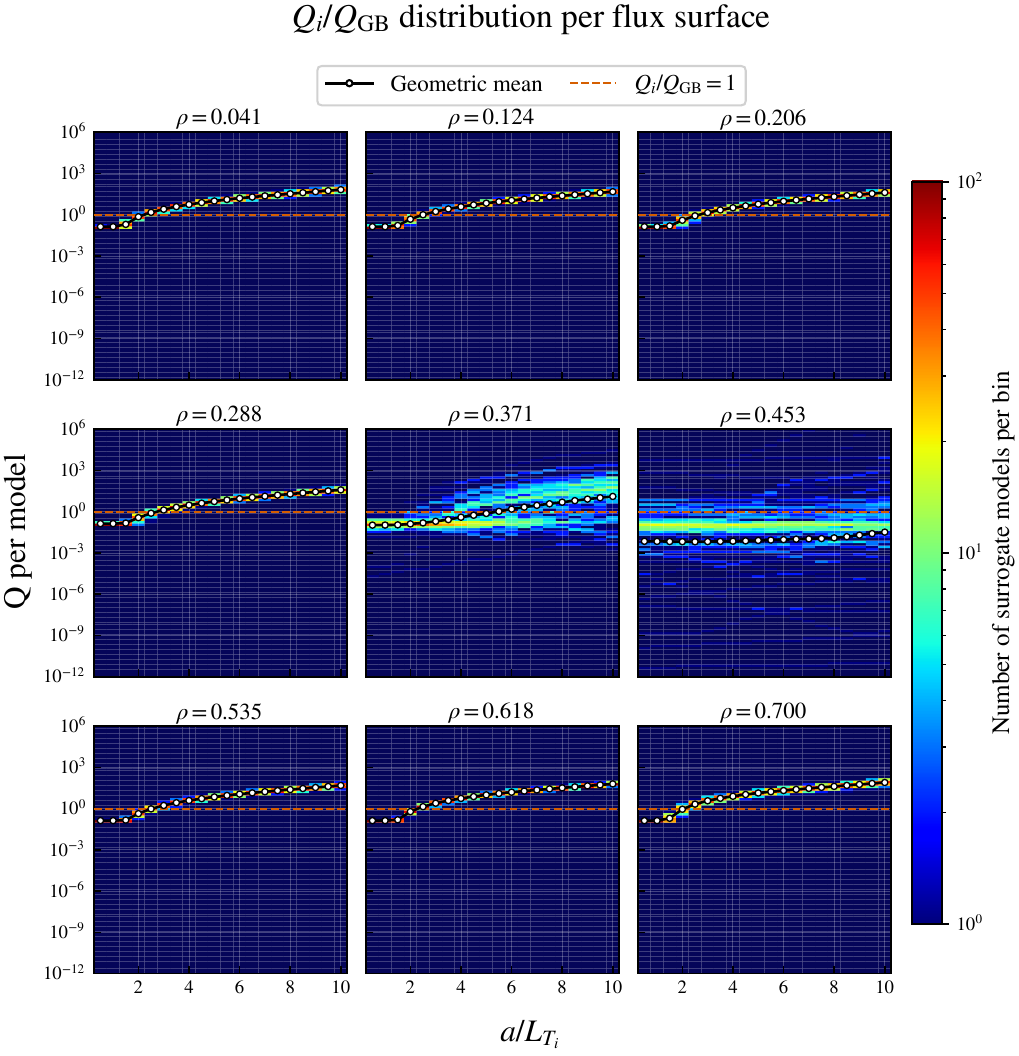}
    \caption{2D histogram of ensemble neural network predictions of heat flux $Q$ versus gradient scale length $a/L_{T_i}$. Shows most flux surfaces are tightly predicted, except for $\rho = 0.371$ and $\rho = 0.453$, where there is a wide spread, indicating an issue with the neural network prediction such as out-of-distribution data.}
    \label{fig:dist_critical_gradient}
\end{figure}

At first glance, there is an obvious best scoring equilibrium towards the end of the campaign (marked in white). However, in this optimized equilibrium lies a cautionary tale on the use of AI surrogates in obtaining a turbulence optimized configuration. Upon investigating the resulting optimized critical gradient across the radial profile, in Figure~\ref{fig:bad_critical}, we can see there are two mid-radius points with abnormally high critical gradients, and we can see in the heat flux versus gradient scale length plots that these two mid-radius points have very low heat flux all the way out to the max of the gradient scale lengths considered ($a/L_{T_i}=10$). We look in more detail at the uncertainty of the prediction, looking at the distribution of predictions based on the ensemble nature of the CNN. We see in Figure~\ref{fig:dist_critical_gradient} that for these two flux surfaces ($\rho = 0.371$ and $\rho = 0.453$) with seemingly abnormally high critical gradient, the uncertainty from the ensemble is very large, especially compared to other flux surfaces, indicating the prediction should not be trusted. 

\begin{figure}
    \centering
    \includegraphics[width=0.95\linewidth]{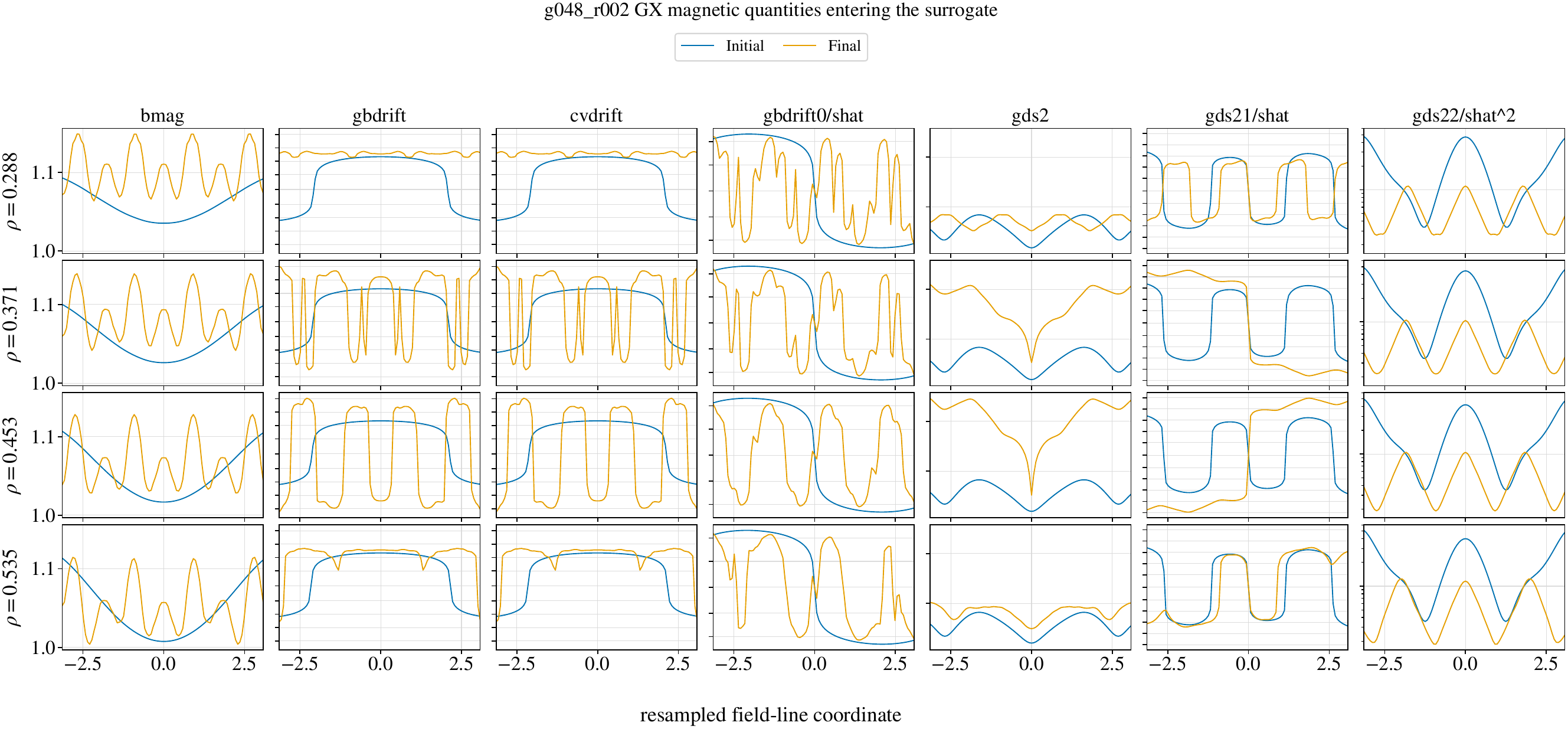}
    \caption{Magnetic quantities for the highest performing equilibrium in the optimization campaign, used by \texttt{GX} and inputs to the AI \texttt{GX} surrogate. This is for four flux surfaces, the two middle being the ones which are suspiciously low heat flux. The differences in e.g., \texttt{gds2} are apparent, and may suggest out-of-distribution from the dataset the surrogate was trained on.}
    \label{fig:mag_quantities}
\end{figure}

The underlying reason for this high uncertainty could be a number of issues, most likely out-of-distribution data. We plot the magnetic geometry quantities which are fundamental to \texttt{GX} and input into the neural network in Figure~\ref{fig:mag_quantities}, for the two problematic flux surfaces and the flux surfaces just before and after. We see even comparing to neighboring flux surfaces and the initial equilibrium, these two problematic surfaces have a number of potential issues which may make it out-of-distribution, notably a very large binormal metric coefficient $\left | \nabla y \right|^2$ (\texttt{gds2}). Further viability checks revealed it fails the nested flux surface check in \texttt{DESC}. It is then not surprising that the AI surrogate failed (and indeed there is the possibility that normal \texttt{GX} would fail also), but it fails in such a way that without proper checks and constraints, it can falsely indicate good performance.     

This issue is beyond the scope of the current work, but we note it as an example of the type of caution that must be exercised with AI surrogates in the optimization loop. We will discuss future work in Section~\ref{sec:future} for detecting out-of-distribution data, for  leveraging multi-fidelity optimization to call out to \texttt{GX} when uncertainty is high, and for AI architectures which will allow for extrapolating better beyond the training dataset. We also note that the fact that non-nested equilibria were not penalized in the score metric means subsequent generations in the genetic algorithm were biased towards these seemingly high scoring parents. Future work further constraining accepted equilibria may help the optimization campaign find better performing configurations. 

\begin{figure}[bht]
    \centering
    \includegraphics[width=0.75\linewidth]{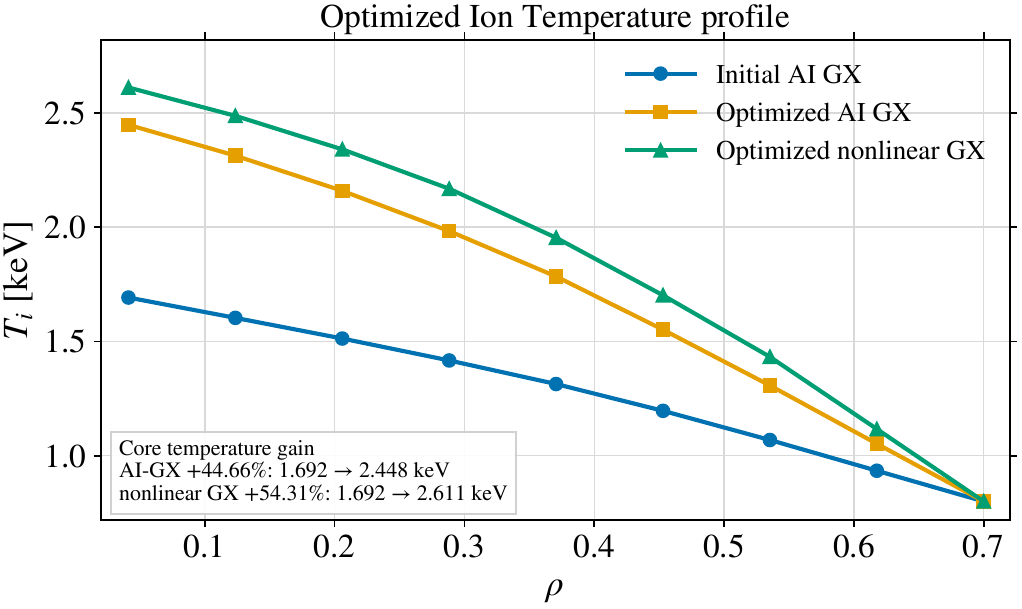}
    \caption{Predictions of the ion temperature using \texttt{T3D} with the best-performing physically valid turbulence optimized equilibrium from the \texttt{autoresearch} runs.}
    \label{fig:best_temperature}
\end{figure}

We investigate the other optimized equilibria, which do pass various tests for good equilibrium, including nested flux surfaces, and run \texttt{T3D} simulations (see Section~\ref{sec:t3d}) to determine which have the most improved core temperature. The median is a ${\sim}$20\% improvement, a modest increase from the 10\% found in previous works using only fixed gradient, single flux tube simulations~\cite{kim_optimization_2024}, and the max is a ${\sim}$44\% improvement, shown in Figure~\ref{fig:best_temperature}. We note also these results can be greatly improved with a surrogate which would allow density and electron temperature evolution also, without having to assume fixed quantities which may not be realistic or achievable. We will further discuss plans for this in Section~\ref{sec:future}.

\subsection{Transport solver acceleration with AI GX surrogate} \label{sec:t3d}
\texttt{T3D}\cite{qian_stellarator_2022} is a transport solver designed for stellarators. It uses the \texttt{TRINITY} algorithm~\cite{barnes_direct_2010} for scale separation in space and time of the transport equations, solving for slower transport timescale evolution using time-averaged, flux tube localized fluxes from faster turbulent transport solvers. These fluid moment transport equations that \texttt{T3D} solves are:

\begin{equation}
\begin{aligned}
\frac{\partial n_s}{\partial t} +
\frac{1}{V'}
\frac{\partial}{\partial \psi}
\left(
V'\langle \overline{\Gamma_s} \rangle
\right)
&=
 \langle \overline{S_{n,s}}  \rangle 
\\[0.5em]
\frac{3}{2}\frac{\partial p_s}{\partial t}
+
\frac{1}{V'}
\frac{\partial}{\partial \psi}
\left(
V' \langle  \overline{Q_s} \rangle
\right)
&=
-\langle \overline{H_s} \rangle  + \frac{3}{2}n_s
\sum_u
\nu_{su}
\left(T_u-T_s\right)
+
 \langle  \overline{S_{p,s}}  \rangle 
\end{aligned}
\label{eq:particle_energy_transport}
\end{equation}

Here, $p_s$, $n_s$, and $T_s$ denote the pressure, number density, and temperature of species $s$, respectively; $t$ is time, $\psi$ is the radial flux coordinate, and $V'=\mathrm{d}V/\mathrm{d}\psi$ is the differential volume associated with that coordinate. The quantity $Q_s$ is the radial heat flux of species $s$, while $H_s$ represents the corresponding volumetric heating or energy-transfer term. Angle brackets denote a flux-surface average, and the overbar denotes the additional averaging applied to the underlying transport quantities, such as a temporal or turbulent average. The summation over $u$ accounts for collisional energy exchange between species $s$ and all other species $u$, with $\nu_{su}^{\varepsilon}$ the associated energy-exchange frequency and $T_u-T_s$ the interspecies temperature difference. Finally, $S_p$ denotes an external or prescribed pressure-source term.

\begin{figure}[bht]
    \centering
    \includegraphics[width=0.75\linewidth]{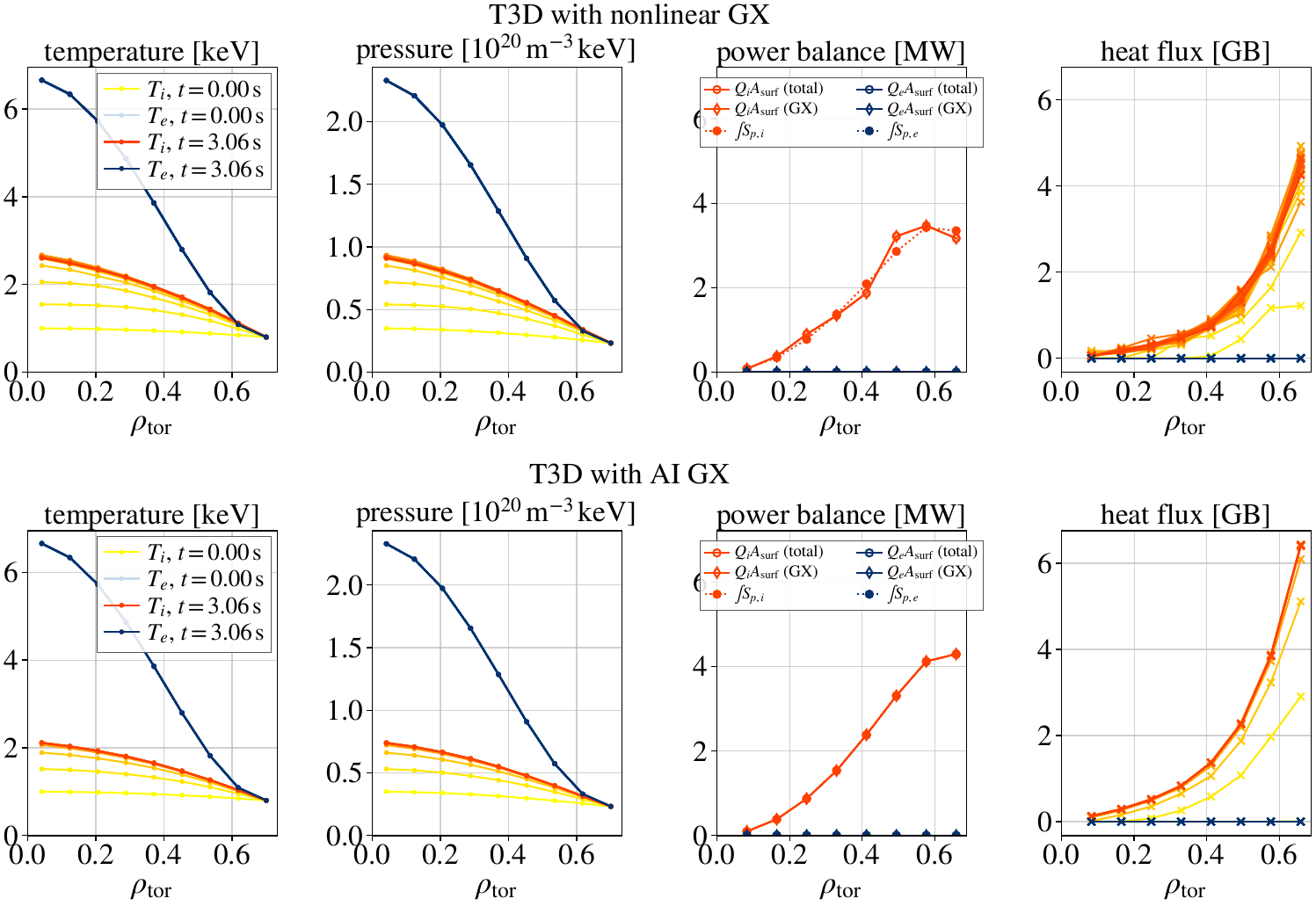}
    \caption{Ion temperature transport evolution from the \texttt{T3D} code. The top row is the simulation using normal \texttt{GX} simulations, totaling 3 hours wall clock time with 16 GPUs. The bottom row is the same simulation but using the AI \texttt{GX} surrogate, totaling 4 minutes using a single GPU.}
    \label{fig:t3d_results}
\end{figure}

The AI \texttt{GX} surrogate was integrated into the \texttt{T3D} code for fast calculations of the turbulence-driven ion heat flux term $\langle \overline{Q_i} \rangle$. As mentioned previously, this limits to only evolving the ion temperature $T_i$, with other plasma state quantities held fixed (e.g., density $n$, electron temperature $T_e$). We solve for $T_i$ using a test case from \texttt{T3D} \cite{beurskens_ion_2021, t3d_w7x_gx} but with the optimized equilibrium from Figure \ref{fig:best_temperature} and compare the results when using normal \texttt{GX} versus using the AI \texttt{GX} surrogate in \texttt{T3D}. This test does not have external heating sources, but rather since initial $T_e \gg T_i$, collisional energy exchange heats the ions. Neoclassical radial transport is not included. The magnetic equilibrium is held constant during the simulation. Nine radial points, each with a single flux tube at $\alpha=0$, is used. The nonlinear GX calculations used $N_\theta=64$, $N_x=N_y=64$, $N_{\mathrm{Hermite}}=12$, and $N_{\mathrm{Laguerre}}=4$, differing from the training-dataset resolution of \((96,64,64,8,4)\) primarily in the parallel and Hermite resolutions. $T_e$ is initialized close to the experimental value, core of 6.7 keV, much higher than $T_i$, which is core of 1.0 keV. Edge density boundary condition is $2.9 \times 10^{19} m^{-3}$, the edge temperature boundary condition is $T_{e,edge} = T_{i,edge} = 0.8$ keV. As can be seen in Figure~\ref{fig:t3d_results}, the final ion temperature from \texttt{T3D} using the normal \texttt{GX} and AI \texttt{GX} surrogate agrees reasonably well. An important point to note is the difference most readily seen in the heat flux at the beginning of the simulation. As \texttt{T3D} begins, the normal \texttt{GX} version has very low flux ($<10^{-6} Q_{GB}$), and steadily increases to its final value. For the \texttt{T3D} run with the AI \texttt{GX} surrogate, the ion heat flux already begins at the first $t = 0$ step with values around $0.1 Q_{GB}$. The evolved ion heat flux converge to approximately the same steady-state value between the two runs, with the inner-most point ${\sim}$10\% lower for the case with AI. We investigated retraining the model with a different objective, removing the earlier limit that clamped heat fluxes to a minimum value of $10^{-2} Q_{GB}$. This improved the lower heat flux predictions at the cost of slightly degraded performance in other regions. This will be further investigated as a future upgrade to the neural network, to ensure that in sequential algorithms like \texttt{T3D} the heat flux calculation will be accurate enough for the dynamical system to evolve to the correct final steady-state profiles. 

In terms of computational performance, \texttt{T3D} calls $(N_\rho - 1) \cdot N_\alpha \cdot (N_p + 1)$ \texttt{GX} evaluations per Newton iteration ($N_\rho$ is number of radial flux surface grid points, $N_\alpha$ the number of poloidal angles, and $N_p$ the number of profile types (e.g., $n$, $T_i$)). The $-1$ is due to mid-grid for flux calculations, and the $+1$ is due to calculating gradients of the flux via profile perturbations (see original \texttt{TRINITY} algorithm~\cite{barnes_direct_2010}). For these simulations, $N_\rho = 9$, $N_\alpha = 1$, $N_p = 1$ and total 31 Newton iterations, for a total of 496 \texttt{GX} evaluations. \texttt{GX} runs can be parallelized over multiple GPUs in space, but must run serially for time evolution. For the AI \texttt{GX} surrogate, it can batch all spatial points and calculate at once. The normal \texttt{T3D} run with \texttt{GX} ran for about 3 hours on 16 GPUs, while the \texttt{T3D} run with the AI \texttt{GX} surrogate ran for 4 minutes on a single GPU, resulting in roughly a 45x wall-clock speedup and 720x GPU-hours speedup.

\section{Limitations and Future Work}\label{sec:future}

This work demonstrates the promise of embedding AI surrogates in optimization loops, enabling direct turbulence objectives and setting the stage for broader integrated physics targets.A current limitation is that the surrogate is trained only on electrostatic, adiabatic-electron simulations. Consequently, the present optimized configurations should be interpreted as optima for electrostatic, adiabatic-electron ITG transport rather than quantitative optima for reactor or experimental performance. Extending to kinetic electrons would enable consistent ion and electron transport solves and optimization toward fusion power, but at dramatically higher cost. Kinetic electron runs are an order of magnitude more expensive, and a naive repetition of the present dataset with longer simulation times and expanded input space could require on the order of 200 million node-hours on exascale systems. Rather than purely learning a static map from geometry to flux, future work may benefit from incorporating more of the underlying dynamics. For example, learning operators or temporal evolution, and richer representations of the distribution function expand the data volume sharply upward. By learning the underlying dynamics these surrogates can potentially help with the challenge of extrapolating beyond the training dataset. These surrogates of the kinetic electron \texttt{GX} simulations would also enable rapid transport solves for experimental planning and interpretation since covering both density, electron temperature, and ion temperature. Additionally it would facilitate, when coupled to an equilibrium solver, self-consistent evolution of the MHD equilibrium and the transport profiles during the optimization. Since neural networks are naturally differentiable, they pair well with differentiable transport solvers, opening the door to gradient-based scenario design and profile control. 

A variety of investigations for improved use of AI surrogates in the optimization loop will be pursued. As seen in Section~\ref{sec:opt_agent}, equilibrium which are too far from the training dataset may fail and may mislead the optimizer. One easy addition would be to include the uncertainty prediction in the optimization penalty function, to steer the equilibrium optimization only to well defined region. A variety of engineering methods for preventing such occurrences through checks on equilibrium viability are good, but do not solve the OOD problem more generally. We will investigate ways for automatic OOD detection and potential continual learning methods for updating the CNN with new simulations that are run. We are investigating multi-fidelity methods in the optimization loop, where for highly uncertain heat flux predictions from the AI surrogate, actual \texttt{GX} simulations are spun up as a higher-fidelity calculation. These simulations if not too many can be run in the simulation loop, and their data retained for further training. We may also be able to employ a reduced, latent representation of equilibrium to be able to detect ``anomalous'' out of distribution data. Additionally, with the speed of the \texttt{T3D} transport solver with the AI-based surrogate, we can incorporate actual ion temperature into the optimization loop, selectively running for certain iterations and utilizing algorithms for gradient-free optimization.

\section{Conclusions}\label{sec:conclusions}
We have demonstrated the utility of an AI \texttt{GX} surrogate in the optimization of stellarator equilibrium and plasma transport models. Not only does their speed allow faster calculations of optimization (72x faster) and transport simulations (45x wall-clock speedup and 720x reduction in GPU-hours), but also opens up new possibilities such as optimizing the critical gradient across the plasma flux surfaces that simply were infeasible previously due to computational cost. We also demonstrate potential pitfalls in these AI surrogates and suggest future work to ensure their successful application to stellarator equilibrium optimization generally, namely: extending to kinetic electron simulations for more complete plasma state, leveraging multi-fidelity optimization techniques for when the AI-based surrogate is uncertain, and continually updating the model. In general AI based surrogates are a promising direction to expand in stellarator optimizations, to bring more direct physics quantities of interest into the optimization, to discover novel stellarator configurations. 

\section*{Acknowledgments}
This work was supported in part by the U.S.~Department of Energy, Office of Science, Office of Advanced Scientific Computing Research and Office of Fusion, Scientific Discovery through Advanced Computing (SciDAC) Program through the FASTMath Institute and the StellFoundry: High-fidelity Digital Models for Fusion Pilot Plant Design Partnership Project under Contract Nos.~DE-AC02-09CH11466 and DE-AC02-06CH11357.
This research uses resources of the National Energy Research Scientific Computing Center (NERSC), a Department of Energy Office of Science User Facility using NERSC awards FES-m4505.

\paragraph{Data}{
Code is available on \url{https://github.com/stellfoundry/ti_optimization}. Data is available on Princeton Data Commons.}

\printbibliography

\appendix

\section{\texttt{autoresearch Codex} prompts}\label{sec:appendix_codex_prompts}
This prompt given to a \texttt{Codex} chat created the instructions for the original attempt at using \texttt{autoresearch} style AI agentic optimization in \texttt{Codex}. The resulting markdown file can be found in the github repository \cite{churchill_ti_optimization} file \texttt{AUTORESEARCH\_GOAL.md}.
\begin{codexprompt}
With ti\_optimization, I have a good setup for phase2 and phase3 for optimizing with the turbulence AI GX surrogate. I want to setup an autoresearch loop, where it tries out different setups for optimization targets and constraints, and also Fourier continuation (first optimizing the first few modes, then freezing and optimized the next). Please write a Markdown that I can give to codex /goal so I can do this. Assume I can run an optimization on a single GPU, and that I will have a node of 4 gpus. The optimization takes 5-10 min each usually. Make sure you also have code to generate an HTML which shows the progress of the different generations, and the top one, and shows details about the different generations.
\end{codexprompt}

This file detailed genetic algorithms and the code to create to run the optimization. It was run in \texttt{Codex} with the prompt \texttt{/goal Run AUTORESEARCH\_GOAL.md}. Note that there is non-determinism is how an LLM will respond, but this prompt shows both what was done and hopefully as an example that can be used and extended.

\section{\texttt{autoresearch} Search Space}
This table gives the search space defined by Codex, used in the equilibrium optimization

\begin{table}[hbt!]
\centering
\caption{Hyperparameter search space used for the
\texttt{autoresearch} genetic algorithm.}
\label{tab:autoresearch_search_space}
\begin{tabular}{ll}
\hline
Parameter & Search space \\
\hline
Fourier continuation
    & $[1,2]$, $[1,2,3]$, $[1,2,3,4]$ \\

$\lambda_{\rm QS}$
    & log-uniform $[0.01,\,0.12]$ \\

$\lambda_{\rm crit}$
    & log-uniform $[0.05,\,1.0]$ \\

$A_{\rm target}$
    & $\{6,\,7\}$ \\

$\lambda_A$
    & log-uniform $[0.03,\,0.5]$ \\

Initial trust radius
    & log-uniform $[0.05,\,0.4]$ \\

$N_{\rm iter,max}$
    & $\{500,\,750,\,1000\}$ \\

$f_{\rm tol},\,x_{\rm tol},\,g_{\rm tol}$
    & log-uniform $[10^{-5},\,10^{-3}]$ \\

$\rho$
    & $\{0.124,0.288,0.453,0.618\}$ or 9-point grid \\

$\alpha$
    & $\{0\}$, $\{0,\pi/2\}$, or
      $\{0,\pi/4,\pi/2,3\pi/4\}$ \\

$a/L_n$
    & uniform $[0.72,\,1.15]$ \\

$a/L_{T_i}$ scan
    & 12 points on $[0.5,6]$ or 20 points on $[0.5,10]$ \\

$Q_{\rm crit}$
    & log-uniform $[0.9,\,1.6]$ \\

$\Delta \log Q$
    & uniform $[0.25,\,0.6]$ \\

$\alpha$ aggregation
    & worst-field-line value \\

\hline
\end{tabular}
\end{table}

\end{document}